\documentclass{elsart}
\usepackage{epsf}

\usepackage{amsmath}
\usepackage{amsfonts}
\usepackage[dvips]{graphicx}

\newcommand{\sci}  [2]{\mbox{$#1 \cdot 10^{#2}$}}

\DeclareMathOperator{\I}   {i}

\newcommand{\E}{\mathrm{e}}
\newcommand{\abs}  [1]{\lvert#1\rvert}

\newcommand{\eqn}  [1]{\mbox{Eq.\! (\ref{#1})}}

\newcommand{\fig}  [1]{\mbox{Fig.~\ref{#1}}}

\newcommand{\figa} [2]{\mbox{Fig.~\ref{#1}#2}}

\newcommand{\sect} [1]{\mbox{Section~\ref{#1}}}

\begin{document}
\begin{frontmatter}
  \title{Composite ``zigzag'' structures in the 1D complex Ginzburg-Landau
    equation} 
  \author[fhi]{Mads Ipsen} and 
  \author[nbi,mpi]{Martin van Hecke}
  \address[fhi]{Fritz-Haber-Institut der Max-Planck-Gesellschaft, Faradayweg
    4-6, D-14195 Berlin, Germany} 
  \address[nbi]{Center for Chaos and Turbulence
    Studies, The Niels Bohr Institute,\\ Blegdamsvej 17, 2100 Copenhagen \O,
    Denmark} 
  \address[mpi]{Max-Planck-Institut f\"ur Physik komplexer Systeme,
    N\"othnitzer Stra{\ss}e 38, D-01187 Dresden, Germany}
  
\begin{abstract}
  We study the dynamics of the one-dimensional complex Ginzburg Landau
  equation (CGLE) in the regime where holes and defects organize
  themselves into composite superstructures which we call {\em
  zigzags}. Extensive numerical simulations of the CGLE reveal a wide
  range of dynamical zigzag behavior which we summarize in a ``phase
  diagram''.  We have performed a numerical linear stability and
  bifurcation analysis of regular zigzag structures which reveals that
  traveling zigzags bifurcate from stationary zigzags via a pitchfork
  bifurcation. This bifurcation changes from supercritical (forward)
  to subcritical (backward) as a function of the CGLE coefficients,
  and we show the relevance of this for the ``phase diagram''. Our
  findings indicate that in the zigzag parameter regime of the CGLE,
  the transition between defect-rich and defect-poor states is
  governed by bifurcations of the zigzag structures.
  
  \noindent{\it PACS:}  
  05.45.Jn, 
  47.54.+r, 
  05.45.Pq 

\end{abstract}
\end{frontmatter}

\section{Introduction}\label{s_intro}

Many extended systems display the formation of patterns when
driven sufficiently far from equilibrium. In the simplest
case, such patterns are regular, but often the patterned state
displays disorder in space and time; this phenomenon is
referred to as {\em spatiotemporal} or {\em extended} chaos
\cite{ch,saka1,saka2,chao1,phasediagram,chate,egolf,clon,mm,maw,review,saar1}.
The tools to describe chaos in low dimensional systems are
often impracticable for the description of extended chaos. In
addition, they are not suited for describing the spatial
organization that takes place in chaotic states that appear to
be built up from {\em local structures}, almost particle-like
entities with well defined dynamics and interactions
\cite{chate}.

The one-dimensional complex Ginzburg Landau equation
\begin{equation}
\partial_t A = A + (1+ i c_1) \partial_{xx} A - (1- i c_3) |A|^2 A~,
\label{cgle}
\end{equation}
 describes pattern formation near a supercritical Hopf bifurcation
\cite{ch} and provides a convenient framework for the study of
spatiotemporal chaos and the role of local structures
\cite{ch,saka1,saka2,chao1,phasediagram,chate,egolf,clon,mm,maw,review,saar1}.
In the CGLE, a wide range of dynamical states can be reached depending
on the coefficients $c_1$ and $c_3$ and the initial conditions: {\em
(i)} For small $c_1$ and $c_3$, plane waves where $A\!\sim\!\exp(i( q
x- \omega t))$ are the dynamically relevant states.  {\em (ii)} For
coefficients $c_1$ and $c_3$ beyond the Benjamin-Feir-Newell curve
where $c_1 c_3 \!=\! 1$, all plane waves are linearly unstable and
spatiotemporally chaotic states occur
\cite{ch,saka1,saka2,chao1,phasediagram,review}. When one takes
sufficiently smooth initial conditions and fixes $c_1$ and $c_3$ just
beyond the Benjamin-Feir-Newell instability, so-called {\em phase
chaos} occurs.  In such states, the value of $|A|$ remains close to
unity and the essential dynamics occurs in the complex phase $\phi$ of
the order parameter $A$. {\em (iii)} For different initial conditions
or larger values of $c_1$ and $c_3$, $A$ can go through zero. At such
points the phase gradient diverges and the phase field shows
topological defects; chaotic states where this happens are referred to
as {\em defect chaos}.  The transition from phase to defect chaos can
either be continuous or hysteretic
\cite{chao1,phasediagram,egolf,maw}. Once defects are formed in the
hysteretic regime, defect chaos persists down to the so-called $L_2$
transition~\cite{phasediagram}.

\begin{figure}
  \vspace{-0cm} \epsfxsize=1.\hsize \mbox{\hspace*{-.01 \hsize}
    \epsffile{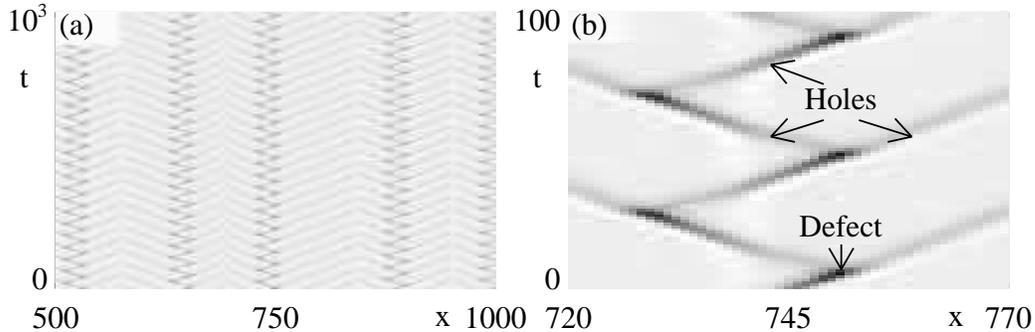} }
  \vspace{-5.8cm}
  \caption{%
    Grey scale (black corresponding to $A\!=\!0$) space--time
    plots of $|A|$ showing basic zigzag phenomenology.  (a) A
    part of a ``frozen'' state of many stationary zigzags
    obtained for $c_1\!=\!0.9$ and $c_3\!=\!1.15$, in a system
    of size 2048, starting from random initial conditions. A
    transient of $t\!=\!10^4$ has been removed. (b) Close--up
    of the core region of a stationary zigzag in (a),
    identifying the defects and holes. }
  \label{fig:ZZPhen}
\end{figure}

Near this $L_2$ transition, spatiotemporally chaotic states
are often built up from {\em holes} (propagating local
concentrations of the phase-gradient) and {\em defects}
\cite{clon,mm}. In this paper we will study the dynamical
states that occur when these holes and defects organize into
more complex composite structures called {\em zigzags}
\cite{clon} (see Fig.~\ref{fig:ZZPhen}). Zigzags consist of a
core where holes alternatingly propagate left and right; in
this process holes are periodically emitted from the core.
Extensive numerical simulations of the CGLE, presented below,
reveal that zigzags display a wide variety of dynamical
behaviors. In addition, while most chaotic states of the CGLE
are characterized by short-time correlations \cite{chao1}, the
relevant time scales of some of the chaotic zigzag states can
be of order $10^3$ or larger (see Fig.~\ref{f3a} and
\ref{f3b}).  We develop numerical methods to perform linear
stability and bifurcation analysis for these complex
structures; these tools are applicable to a wide range of
complex local structures such as oscillating sources
\cite{source}, oscillating domain walls \cite{dw} and
oscillating pulses \cite{pulse}. The results of our analysis
indicate that, in the zigzag dominated regime, the $L_2$
transition is closely related to local bifurcations of the
zigzags\footnote{A similar conclusion relating bifurcations to
the $L_1$ and $L_3$ transitions was formulated in \cite{maw}}.

The paper is organized in two parts. We first study the
phenomenology of zigzags, and discuss the role of the holes
and defects that constitute their building blocks (section
\ref{pheno}).  Then we perform a numerically very demanding
linear stability and bifurcation analysis of zigzags (section
\ref{spec})), and show how some of the main features of the
zigzag phenomenology can be understood from this analysis.

\section{Phenomenology of zigzags}\label{pheno}

The zigzag structures display a wide variety of dynamical behaviors as
a function of the coefficients $c_1$ and $c_3$ of the CGLE (see
Figs.~\ref{f3}--\ref{f3b}). In section \ref{zz1} we will discuss the
local structures that form the zigzags and discuss various types of
regular zigzags. An overview of regular and irregular zigzag dynamics
in large systems and for long integration times is presented in
section \ref{large}.

\subsection{Ingredients of zigzags: holes and defects} 
\label{zz1}

Some of the qualitative properties of zigzags can be
understood from the properties of the homoclinic holes and
defects that are the building blocks of zigzags and that have
been studied in a series of recent papers
\cite{clon,mm,maw}. The main ingredients of importance here
are briefly summarized below. {\em{(i)}} Coherent homoclinic
holes are localized packets of phase gradient that have the
special property that they can propagate uniformly, i.e., they
are of the form $A(x,t)\!=\!  \exp{i \omega t} \bar{A}(x-vt)$
\cite{clon}. Right (left) moving holes are characterized by a
positive (negative) phase gradient peak in their
core. {\em{(ii)}} Coherent holes are linearly unstable, and in
phase space they form a saddle with a one-dimensional unstable
manifold \cite{clon}. When they are perturbed so-called {\em
incoherent} holes are observed, i.e., holes that evolve over
time; in phase space we can think of these holes as evolving
along the one-dimensional unstable manifold of the coherent
holes. Incoherent holes finally either decay or evolve towards
defects, and the closer an initial condition is towards the
stable manifold of the coherent holes, the longer its
lifetime.  {\em{(iii)}} The spatial profile that occurs
shortly after a defect has occurred consists of a
juxtaposition of a negative and positive phase gradient
peak. The negative peak can initiate a left moving hole, and
the positive peak a right moving hole.

\begin{figure}
  \vspace{0cm} \epsfxsize=1.\hsize \mbox{\hspace*{-.01\hsize}
    \epsffile{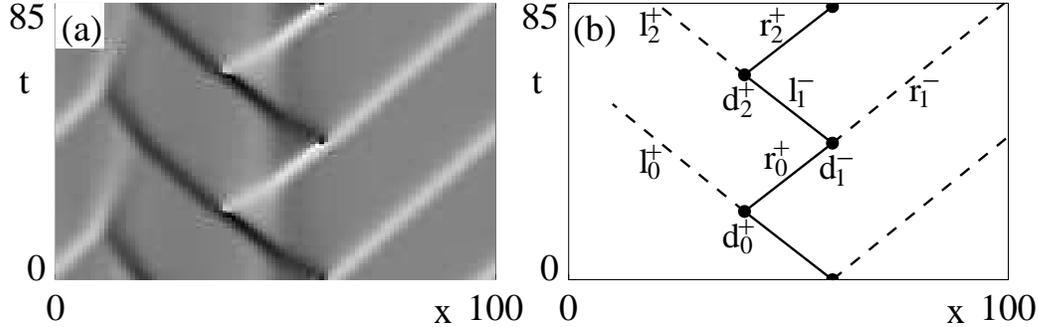} } \vspace{-5.8cm}
\caption[] {(a) Grey scale plot for the phase gradient for a stationary
zigzag. White (black) corresponds to positive (negative)
phase gradient; in the grey areas the phase gradient is close to
zero. (b) Schematic structure of the same zigzag as in (a).  Defects
are labeled with a $d^\pm$, where the sign reflects the change in
winding number $\int dx \partial_x \phi$ before and after their
occurrence; after the generation of a $d^+$ ($d^-$) defect the winding
number has increased (decreased) by $2 \pi$.  The holes are labeled
$(l,r)$ depending on their direction of propagation, and are created
by and evolve to defects. Holes carry the sign of the defect that
generated them as an additional label. A left (right) moving hole can
only generate $d^+$ ($d^-$) defects \cite{clon,mm}.  }\label{f2}
\end{figure}

The traveling holes that occur in our zigzag states (see
Fig.~\ref{f2}a) are incoherent, and as expected their
direction of propagation is governed by the sign of their
phase gradient (see Fig.~\ref{f2}a).  Let us inspect the
zigzag structure in detail, starting from the defect $d_1^-$
in Fig.~\ref{f2}b. This defect generates two new incoherent
holes, $l_1^-$ and $r_1^-$. The former stays in the core and
rapidly generates another defect $(d_2^+)$, while the latter
is send out of the zigzag core. The difference in lifetimes
between $l_1^-$ and $r_1^-$ is related to the details of the
defect profile of $d_1^-$.  In the regime of the CGLE where
zigzags occur, the positive peak of a $d^-$ defect is closer
to the stable manifold of a right moving hole, than the
negative peak is to the stable manifold of a left moving
hole. As a result, the lifetime of the $r^-$ holes is larger
than that of $l^-$ holes. As far as we understand, this is the
essential condition that produces zigzags.  The left-right
symmetry of the CGLE inverts the sign of phase gradients,
implying that the lifetime of an $r$ hole emanating from a
$d^-$ defect is similar to the lifetime of an $l$ hole
emanating from a $d^+$ defect. This symmetry can be (weakly)
broken, giving rise to drifting zigzags (see below).

\begin{figure}
  \vspace{0.0mm} 
  \epsfxsize=1.\hsize \mbox{\hspace*{. \hsize}
    \epsffile{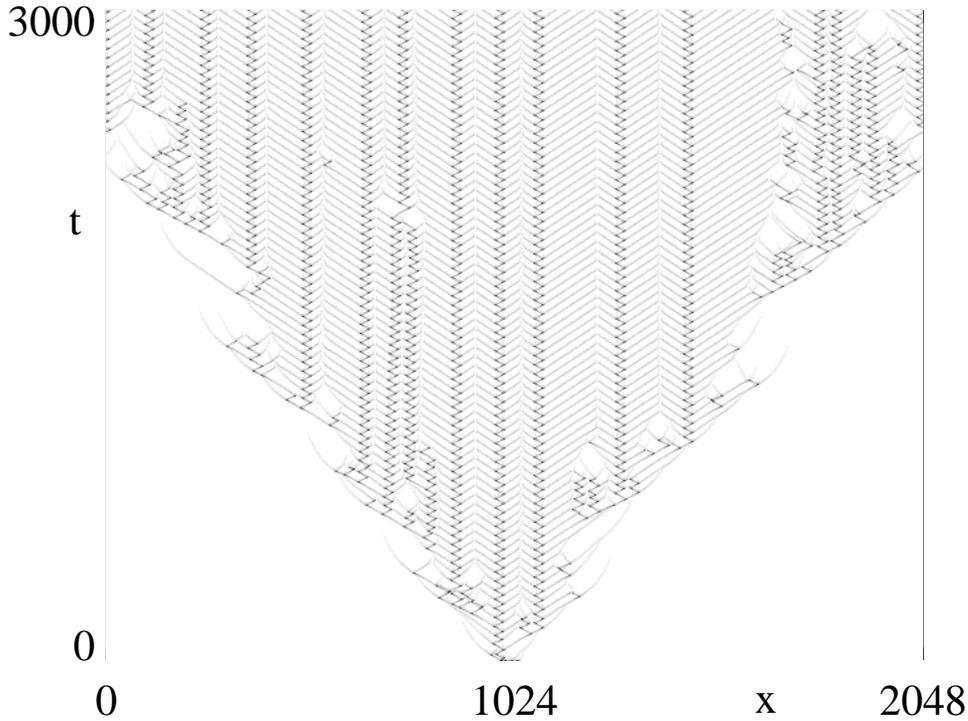} 
    } 
    \vspace{0.0cm}
  \caption{%
    Example of the spreading out of zigzags when the wing holes
    are active, for $ c1\!=\!0.9$ and $c3\!=\!1.15$. }
\label{fig:SpreadingZZ}
\end{figure}

It is important to note that for values of $c_1$ and $c_3$
away from the zigzag regime, hole-defect dynamics with
completely different dynamics may occur. For example, when the
time scales of $r^+ \rightarrow d^-$ and $r^- \rightarrow d^-$
are similar more disordered dynamics as shown in Fig.~1 of
\cite{clon} sets in, while when $r^+ \rightarrow d^-$ is {\em
slower} than $r^- \rightarrow d^-$, the dynamics is dominated
by propagating incoherent holes \cite{mm}.

\subsubsection{Wing holes}
\begin{figure}
\vspace{00mm} \epsfxsize=1.11\hsize \mbox{\hspace*{-.05 \hsize}
\epsffile{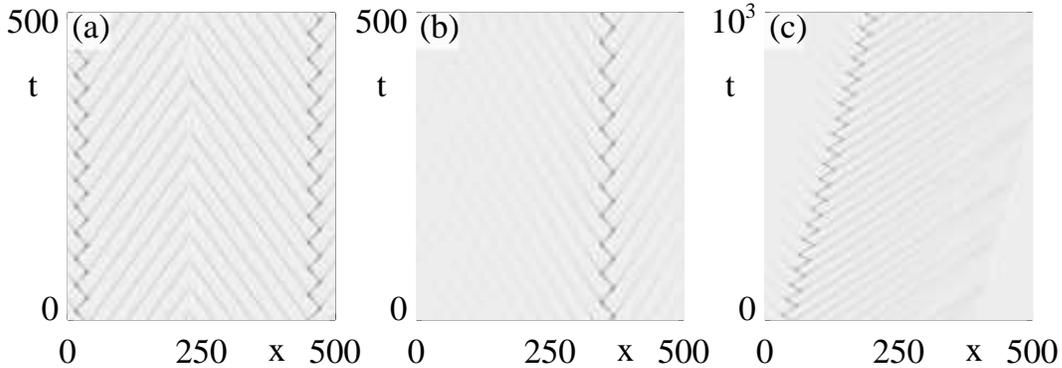} }
\vspace{-6.5cm}
\caption[] {Illustration of the difference between active and decaying
wing holes, and stationary and traveling zigzags. All panels show
$|A|$ from a longer run in a system of size 2048.  (a) Active
wing holes for $c_1\!=0.9$ and $c_3\!=\!1.14$. Here the final states
consist of a number of stationary zigzags, separated by a range of
distances. Here we show the area in between two widely separated
zigzags. For this spacing the wing holes are very close to being
critical and for larger separation they would grow out to
defects. Some effect of this is seen in the fluctuation near
$x\!=\!250$ and $t\!=400$.  (b) For $c_1\!=\!0.8$ and $c_3\!=\!1.13$
the wing holes clearly decay, and zigzags are essentially
isolated. (c) The late dynamics in a run for $c_1\!=\!0.6$ and
$c_3\!=\!1.2$, showing an isolated traveling zigzag.}\label{f3}
\end{figure}

Holes that are send out of the zigzag core (like hole
solutions indicated by $l_2^+$ and $r_1^-$ in Fig.~\ref{f2})
are referred to as {\em wing holes}.  For regular zigzags, the
holes a generated periodically, yielding traveling periodic
arrays of holes, with hole-hole spacings ranging from 20-100;
the size of this spacing is typically sufficient for regarding
the holes as isolated.  These wing holes can either evolve to
defects or decay.  In the former case, zigzag dynamics spreads
throughout a laminar background (Fig.~\ref{fig:SpreadingZZ})
and holes collide in shock-like structures that separate
neighboring zigzags (Fig.~\ref{f3}a); in this case the maximal
spacing between zigzags is set by the wing hole lifetime.
When the wing holes decay isolated zigzags can occur
(Fig.~\ref{f3}b).

\subsubsection{Core holes and drift}
A priori one does not know whether the hole-defect composite
that makes up the core of the zigzags is linearly stable,
although the examples shown in Fig.~\ref{fig:ZZPhen} and
\ref{f3} clearly indicate that this can be the case. In some
cases however, stationary zigzags can become linearly
unstable. We will present ample examples of this below, but
here we will already point out the main consequence: depending
on the CGLE coefficients, stable zigzags can either be
stationary (Fig.~\ref{f3}a,b) or drifting (Fig.~\ref{f3}c). In
section \ref{spec} we will show that the drifting zigzags
branch off from the stationary zigzags via a pitchfork
bifurcation.

\subsection{Large scale zigzag dynamics}\label{large}

We studied zigzag behavior on large domains $(L=2048)$ and for long
integration times of the order $10^4$ -- $10^5$, and examples of the
different dynamical states that we obtained are presented here. 

\begin{figure}
\vspace{00mm} \epsfxsize=2.\hsize \mbox{\hspace*{-.1 \hsize}
\epsffile{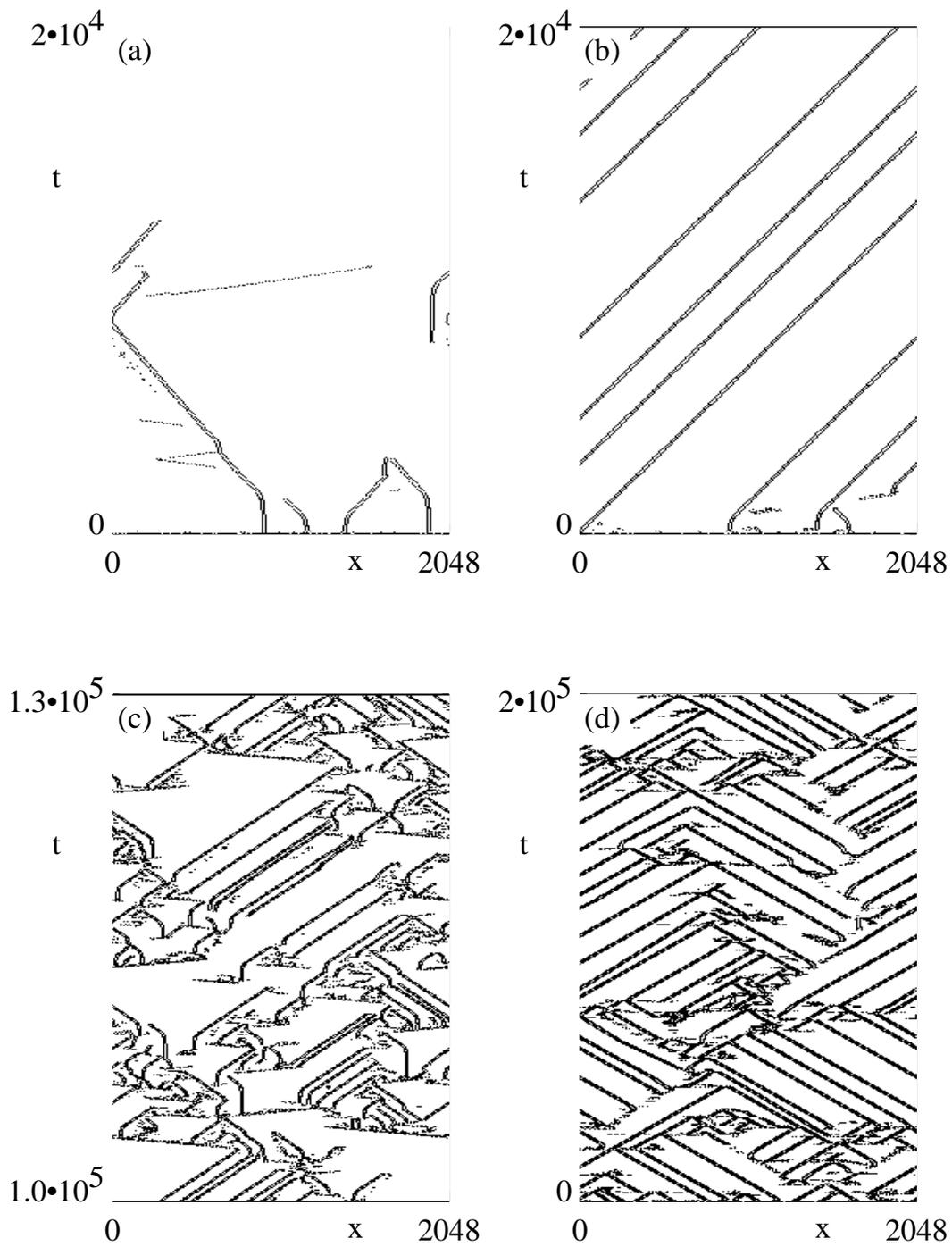} }

\vspace{-0.5cm}
\caption[] {Four examples of zigzag dynamics dominated by drift.  (a)
$c_1\!=\!0.6 $, $c_3\!=\!1.16 $.  (b) $c_1\!=\!0.6 $, $c_3\!=\!1.2 $.  (c)
$c_1\!=\!0.6 $, $c_3\!=\!1.25 $.  (d) $c_1\!=\!1.0 $, $c_3\!=\!1.12 $.  For
more details see text.}\label{f3a}
\end{figure}

\begin{figure}
  \vspace{0.0mm} 
  \epsfxsize=2.\hsize \mbox{\hspace*{-.1 \hsize}
    \epsffile{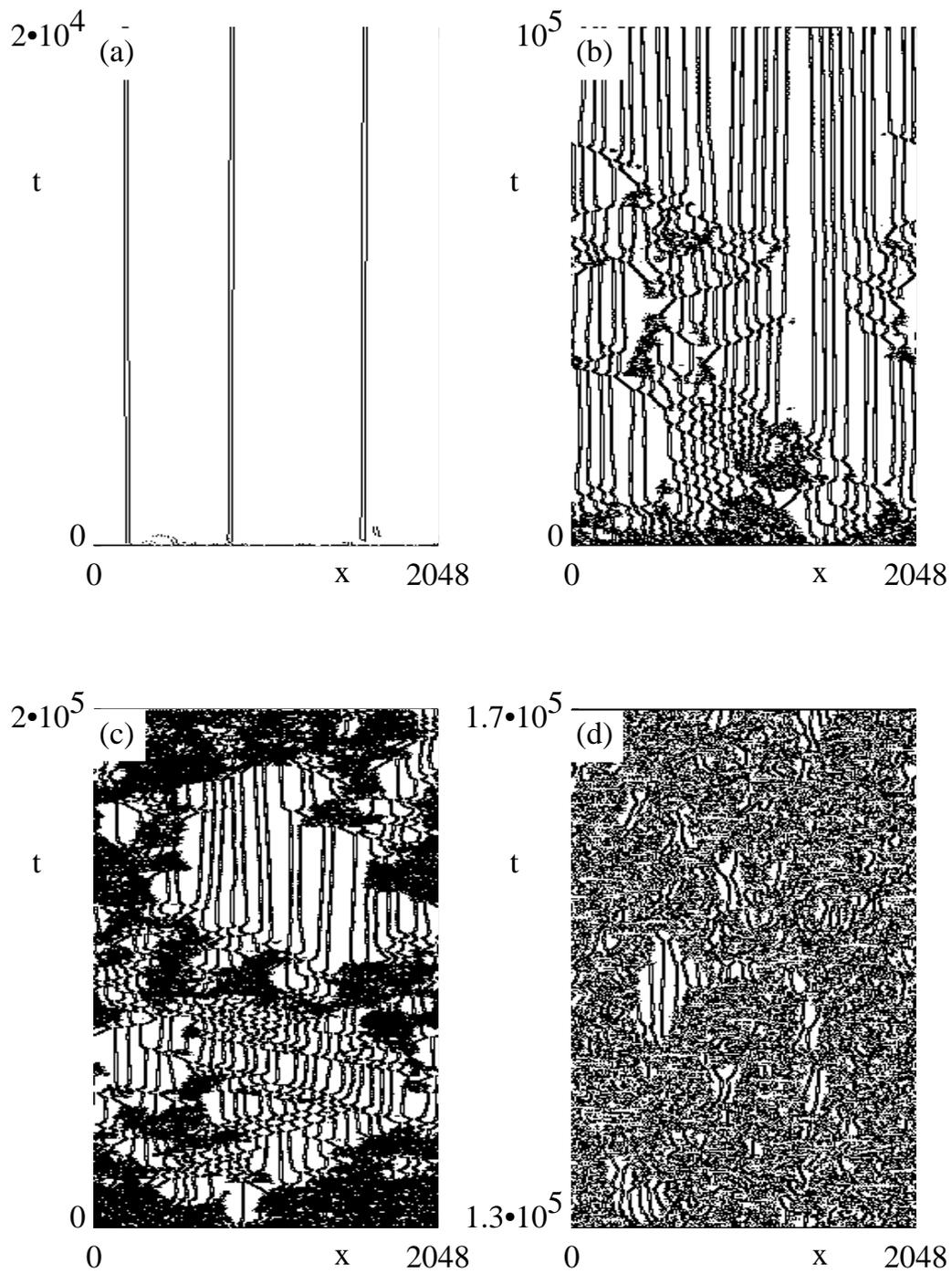} }

  \vspace{-.5cm}
  \caption{Four examples of dynamics dominated by stationary zigzags.  (a)
    $c_1\!=\!0.8 $, $c_3\!=\!1.17 $.  (b) $c_1\!=\!0.9 $,
    $c_3\!=\!1.23 $.  (c) $c_1\!=\!0.9 $, $c_3\!=\!1.24 $.
    (d) $c_1\!=\!0.9 $, $c_3\!=\!1.25 $.  For more details see
    text.}
  \label{f3b}
\end{figure}

Here we will present eight qualitatively different examples of zigzag
dynamics, starting from random initial conditions. The dynamical
states that we will present are characterized by such large spatial
and temporal scales that space-time diagrams of the modulus of $A$,
such as those shown in Fig.~\ref{fig:SpreadingZZ}, become completely
cluttered. We therefore only plot the position of all defects in a
space-time plot (for details regarding the algorithm used for the
detection of defects, see appendix~\ref{app}).  For a single
stationary regular zigzag these defects occur alternatingly on two
spatial positions (see Fig.~\ref{fig:ZZPhen} and~\ref{f2}). On the
time scale shown in Figs.~\ref{f3a} and~\ref{f3b} these individual
defects can no longer be distinguished, and zigzag structures show up
as two parallel lines.

In Fig.~\ref{f3a} we show four examples of dynamics dominated
by traveling zigzags.  Panel (a) shows a long transient that
occurs when $c_1\!=\!0.6$ and $c_3\!=\!1.16$ are just below
the $L_2$ transition. Initially, a few stationary zigzags are
created, but these appear linearly unstable and give rise to
the formation of a few traveling zigzags. These, however, do
not sustain, and after a period of the order $10^4$ the
dynamics decays back to simple uniform oscillations where
$A=\exp{i \omega t}$.  When $c_3$ is increased to a value of
$1.2$, the traveling zigzags become stable, and a state
consisting of homogeneously drifting zigzags occurs
(Fig.~\ref{f3a}b). The positions and overall drift of this
state are selected by the initial conditions. Note that for
early times a left and right moving zigzag collide (around
$x\!=\!1600$) which results in the destruction of the left
drifting zigzag. The large spacing between neighboring zigzags
indicates that the wing holes decay here, similar to
Fig.~\ref{fig:ZZPhen}c.  When $c_3$ is increased even further
to a value of $1.25$, more complicated dynamics occurs
(Fig.~\ref{f3a}c). Left and right drifting zigzags compete,
and intermittent collisions between zigzags take place from
which new zigzags may be created.  In some cases we have
observed that similar states after a very long transient may
decay into a state similar to Fig.~\ref{f3a}b where either
left or right drifting zigzags occur.  Behavior with similar
features occurs for $c_1\!=\!1$ and $ c_3\!=\!1.12$
(Fig.~\ref{f3a}d), although the drift of the zigzags here is
approximately 10 times slower (notice the difference in time
scales between Fig.~\ref{f3a}c and d). It is not clear whether
or not the states in Fig.~\ref{f3a}c--d should be considered
qualitatively the same or not, nor what an appropriate order
parameter for their description should be. This is of course a
general problem that we encounter when we try to classify
behavior as rich as that shown in these figures.

Fig.~\ref{f3b} shows four examples of CGLE dynamics dominated
by stationary zigzags. For $c_1\!=\!0.8$ and $c_3\!=\!1.17$
stationary zigzags occur (Fig.~\ref{f3b}a). The spacing
between adjacent zigzags indicates that the wing holes decay
here. In Fig.~\ref{f3b}b--d three examples of intermittent
zigzag dynamics obtained for $c_1\!=\!0.9$ and $c_3 = 1.23$,
$1.24$, and $1.25$ are shown. The dynamics in Fig.~\ref{f3b}b
displays a disordered transient that decays to a stationary
``glassy'' state of zigzags.  Note that the zigzags themselves
appear as a substrate on which dynamics on even longer space
and time scale occurs; an example is the traveling
perturbation seen around $x\!=\!800$ and $t\!=\! 4 \times
10^4$. This suggest a hierarchy of scales: holes and defects
form zigzags, zigzags form even larger structures, etc.  When
$c_3$ is increased to 1.24, perturbations of the stationary
zigzags do no longer decay, and very disordered dynamics
occurs (Fig.~\ref{f3b}c). Note however, that stationary
zigzags by themselves are not unstable (as evidenced by the
large stationary regime in the middle of this figure), but
only get perturbed by ``contaminations'' coming from chaotic
patches that spread through the system.  This behavior is
typical for spatiotemporal intermittency \cite{phasediagram}.
Finally, when $c_3$ is increased even further to a value of
$1.25$, the disordered patches become much more dominant,
although some pockets filled with stationary zigzags occur
(Fig.\ref{f3b}d).

\subsubsection{Phase diagram}
\begin{figure}
  \begin{center} \includegraphics[scale=0.78]{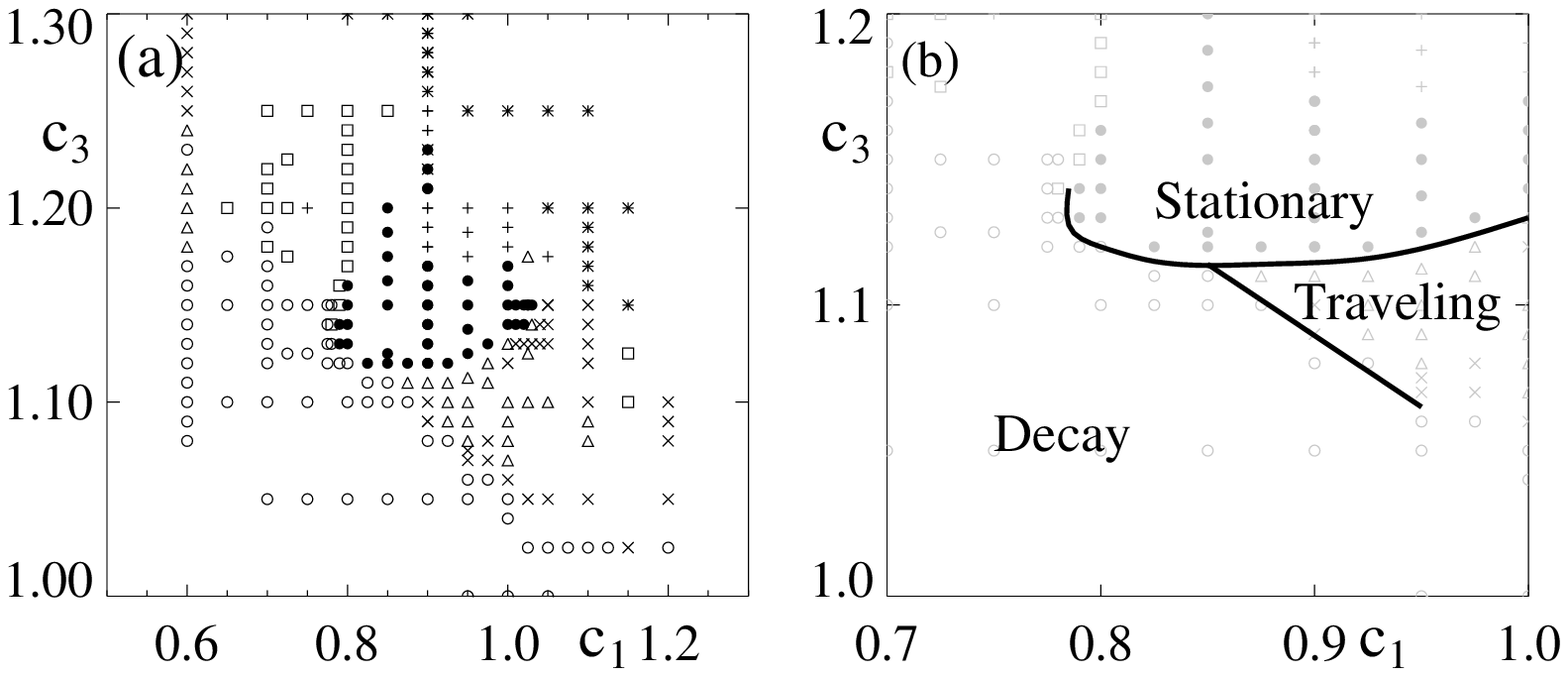} \end{center}
    \caption{(a) Phase diagram showing the variation of the different
    zigzags types found in the $c_1$ and $c_3$ parameter plane. For
    explanation of symbols: see text. (b) Magnification of (a) showing
    the main transitions between stationary, traveling and decaying
    zigzag structures. } \label{fig:PhaseDiagram}
\end{figure}

Based on numerical simulations like the ones presented above,
we have attempted to classify the various types of zigzag
dynamics into a small number of distinct classes; this results
in a ``phase diagram'' for zigzag behavior
(Fig.~\ref{fig:PhaseDiagram}).  This diagram constitutes a
small part of the full phase diagram of the one-dimensional
CGLE only.  The simplest state that can be distinguished is
where, after a transient, all defects disappear (see
Fig.~\ref{f3a}a); coefficients for which this happened in our
simulations are represented in Fig.~\ref{fig:PhaseDiagram} by
an open circle.  States which are dominated by stationary
zigzags (like Fig.~\ref{f3b}a) are represented by a full
circle.  Then there are states which are dominated by
traveling zigzags (Fig.~\ref{f3a}b); these are here
represented by a triangle.  Even when stationary or traveling
zigzags appear to be linearly stable, the overall dynamics can
be disordered. In a sense, these states represent examples of
what one may call ``spatiotemporal intermittency of zigzags'';
such intermittent states with stationary zigzags (like those
in Fig.~\ref{f3b}b--c) are represented by a plus symbol, while
intermittent states with traveling zigzags like
Fig.~\ref{f3a}c--d are represented by a '$\times$'
symbol. When the coefficients $c_1$ and $c_3$ are increased
sufficiently, pure defect chaos ensues (see Fig.~\ref{f3b}d);
these states are represented by a star. Finally, one
occasionally finds states that do not obviously fall into one
of these categories; we have labeled these by boxes.

It should be noted that for all the points in the phase diagram we
performed a single run in a large system ($L = 2048$) and for long
integration times $T\!=\!10^5$, and that each individual run consumes
a considerable amount of (super) computing time.  Instead of trying to
obtain better statistics or a finer spacing of the coefficients for
which we performed runs, we have focussed on what we believe is the
main feature of the bifurcation diagram (see
~\fig{fig:PhaseDiagram}b): {\em(i)} There exists a finite coefficient
regime dominated by stationary zigzags. {\em(ii)} At the bottom
boundary of this regime, one either finds decaying states or traveling
zigzags.  The transition between stationary zigzags and decaying or
traveling zigzags can be understood from the numerical bifurcation
analysis we present in section \ref{spec} below.
 
\section{Linear stability and bifurcation analysis}
\label{spec}

\subsection{Stability}\label{subsec:stab}
We have performed a linear stability analysis of regular
zigzag structures to gain some understanding of their large
scale dynamics.  After the spatial degrees of freedom of the
CGLE are discretized, the problem of finding a regular zigzag
and its spectrum can be translated into finding a periodic
orbit and its spectrum in a set of coupled ODEs. Using
standard continuation algorithms, it is then possible to track
the spectrum as a function of the coefficients $c_1$ or $c_3$,
thus obtaining stability limits and bifurcation points.  This
strategy is straight-forward but numerically demanding. For
low-dimensional systems of autonomous ODEs, such procedures
are standard \cite{Auto97,Kutz95}.  Such analysis is already
well-described in the literature (see e.g.\ \cite{Og2000}),
but has so far mostly been applied to the study of spatial
structures such as uniformly traveling fronts and spots, whose
spatio-temporal dynamics is essentially stationary.  Here we
describe the application of these techniques for structures,
such as the zigzags, which are periodic in time.

\begin{figure}
  \vspace{0cm} \epsfxsize=1.\hsize \mbox{\hspace*{-.01 \hsize}
    \epsffile{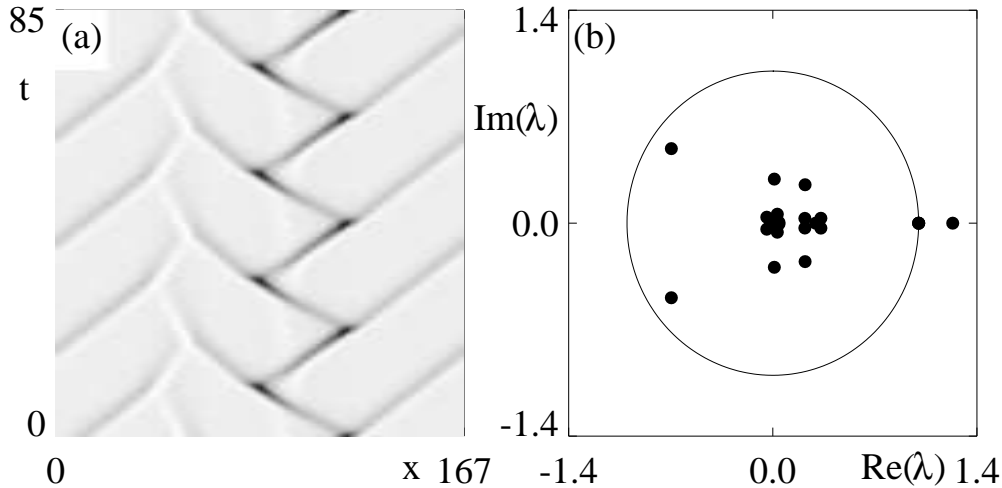} } \vspace{-3.3cm} \caption{ (a) Unstable zigzag
    obtained for $c_1 = 0.9$ and $c_3 = 1.15$ with period $T = 41.64$,
    global phase shift $\phi = -2.80$, and velocity $v = -0.0218$. The
    spectrum associated with this zigzag is shown in (b).  Note the
    neutral eigenvalues at $\lambda=1$ and the single unstable
    eigenvalue.}  \label{fig:TestZZ}
\end{figure}

\paragraph*{Symmetries and boundary conditions}
The choice of the appropriate frame for the CGLE is essential.
To be able to study uniformly drifting zigzag structures we
choose a co-moving coordinate frame. During each period of the
zigzag the phase of $A$ increases by a global phase shift
$\phi$ and the field $A$ is therefore quasiperiodic for a
zigzag; however, by going to a ``rotating'' frame
$\sim\E^{\I\phi t}$, $A$ can be made periodic. Hence we study
the CGLE in the following form:
\begin{equation}
\partial_t A = (1+ i \phi) A + v \partial_x A + (1+ i c_1)
\partial_{xx} A - (1- i c_3) |A|^2 A~,
\label{eq:CgleScaled}
\end{equation}
where $A$ is periodic provided that $\phi$ and $v$ are equal
to the global phase shift and the drift velocity of the
zigzag.  The form of \eqn{eq:CgleScaled} permits us to
determine a zigzag numerically by using a shooting algorithm
similar to the approach applied when determining limit cycle
solutions to autonomous systems of ODEs; for details, see the
appendix~2--3. An example of an unstable zigzag determined by
this method and its spectrum is shown in
Fig.~\ref{fig:TestZZ}.

Due to the periodic boundary conditions wing holes will
collide and annihilate in a shock area (see
Fig.~\ref{fig:TestZZ}). Their role is unimportant when they
are sufficiently far away from the zigzag core, since the
group velocity is pointing towards such shocks and no
``information'' can flow out of them. The position of these
shocks is not arbitrary; for the chosen domain sizes, at most
a number of discrete spacings between the zigzag core and the
shock are available (see Fig.~\ref{f2}b); in general we have
chosen our initial and pinning conditions such that the shocks
are positioned as far from the zigzag core as possible.

Since coherent holes are linearly unstable, one may wonder
whether the corresponding unstable eigenvalues would appear in
the spectrum.  However, the holes that occur in zigzags are
incoherent and have a finite lifetime, because they evolve to
defects (as happens in the core) or are annihilated (as
happens to the wing holes).  It is around these states that
one studies the stability now; the linear stability of the
hole composite zigzag can only be obtained by studying the
full spectrum of the composite structure.  Another example
where linearly unstable coherent holes give rise to partly
regular incoherent hole dynamics can be found in \cite{mm}.

\subsection{Numerical bifurcation analysis of zigzags}
\label{numbif}

In the previous sections, we have already seen that both
steady and traveling zigzags exists as stable solutions to the
CGLE.  Here we discuss how transitions between these two
states can be described using the continuation strategy
discussed in \sect{spec} for determining the stability and
structural properties of a zigzag pattern.

Throughout, we choose $c_1$ and $c_3$ as free parameters, and
focus on the region in the phase diagram dominated by
transitions between regular stationary and traveling
zigzags. This corresponds to the region of the phase diagram
highlighted in~\fig{fig:PhaseDiagram}b; In particular, we
shall focus on the transitions from stationary to either
decaying or traveling zigzags respectively.

\begin{figure}[t]
  \begin{center} \vspace{-0cm} \epsfxsize=1.0\hsize
     \mbox{\hspace*{-.04 \hsize} \epsffile{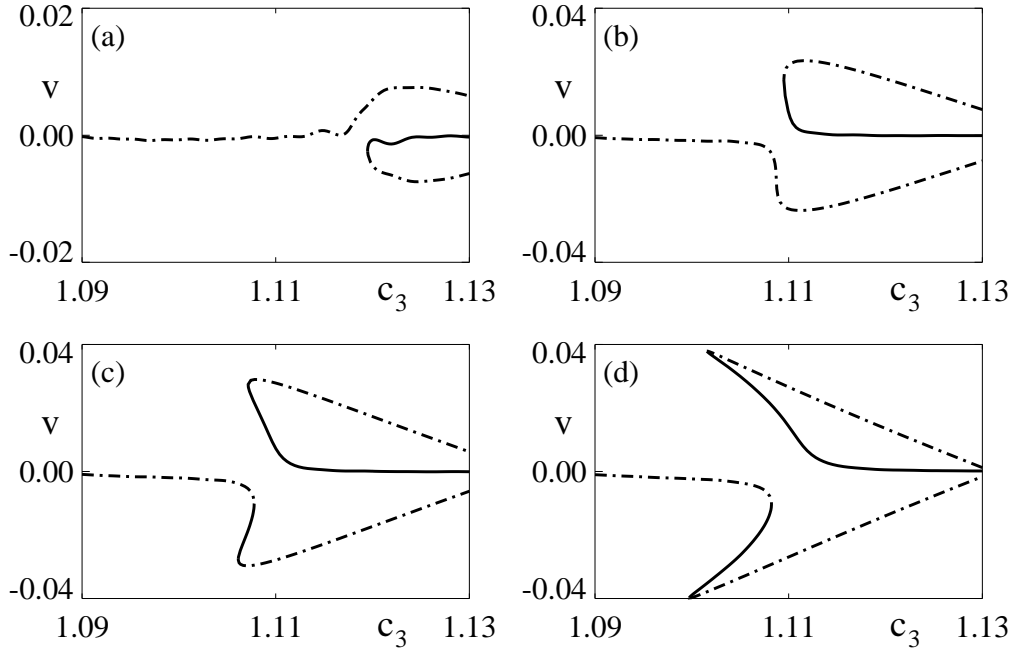} }
        \vspace{-1cm} 
\caption
     { Bifurcation diagrams showing the variation of the zigzag
     velocity $v$ as a function of $c_3$ for $c_1 = 0.80, 0.85, 0.87$,
     and $0.90$ corresponding to (a-d) respectively.  }
     \label{fig:BifSketch} \end{center}
\end{figure}

We have employed the following strategy: for different fixed
values of $c_1$, we make a vertical continuation scan through
the phase diagram in \fig{fig:PhaseDiagram} by varying $c_3$.
The results for $c_1 = 0.80, 0.85, 0.87$, and $0.90$ are shown
in ~\fig{fig:BifSketch}a-d, where the variation of the zigzag
velocity $v$ is shown as a function of $c_3$.  For $c_1 =
0.80$, the stationary zigzag is initially unstable for $c_3 <
1.12$, above which it becomes stabilized via a subcritical
pitchfork bifurcation, which gives birth to two unstable
branches of left and right traveling zigzags.  The perfect
symmetry of the pitchfork is slightly perturbed since we work
on a spatial domain of finite size.  Effectively, this renders
the zigzag slightly asymmetric implying that the bifurcation
diagram exhibits the typical behavior found near a perturbed
pitchfork, where the symmetric ``fork'' splits into two
separate branches containing a fold (saddle-node) point and a
simple unbranched state.  Since this splitting solely is due
to a finite-size effect, we shall refer to the above
bifurcation as a ``pitchfork'' bifurcation. For $c_1 = 0.80$,
the stationary zigzag remains stable for $c_3 < 1.16$, where
it merges with the branches of the traveling zigzags and
looses stability via a second subcritical pitchfork
bifurcation (not shown).  The location of the lower
bifurcation point is in excellent agreement with the
transition from stationary to decaying zigzags observed in the
phase diagram.

Qualitatively, the results for $c_1 = 0.85$ shown
in~\fig{fig:BifSketch}b are similar to the behavior described
above, except that the slopes of the bifurcating branches of
traveling zigzags near the lower pitchfork point have
increased. This suggests that a transition from a sub- to a
supercritical pitchfork may occur as $c_1$ is increased
further.  This is confirmed in~\fig{fig:BifSketch}c, where
$c_1$ is fixed at $0.87$.  Here the branches of traveling
zigzags now branch off from the pitchfork point in the
direction of {\em decreasing} $c_3$.  As $c_3$ decreases
further, a fold point is reached at which both of the
traveling zigzag solutions are destabilized.  Effectively, the
transition from sub- to supercriticality generates a parameter
region in which {\em traveling zigzags are stable}. This
region is localized below the region where stable stationary
zigzags exist. The change of the bifurcation from sub- to
supercritical is in agreement with the emergence of the region
of traveling zigzags in the phase diagram. Finally, for $c_1 =
0.90$ the range of $c_3$ within which stable traveling zigzags
occur increases (see ~\fig{fig:BifSketch}d) in correspondence
with the observation from phase diagram.

\section{Discussion}\label{disc}

In this paper, we have discussed both the structural and
dynamical properties of the family of \emph{zigzag} solutions
of the CGLE.  Some of the structural properties of the zigzags
can be understood in terms of the properties of their
constituent homoclinic holes and defects
\cite{clon,mm,maw}. To understand our finding that zigzag
structures can either be stationary or traveling, we have
employed a stability and bifurcation analysis, which shows
that the two types of zigzags are related by a pitchfork
bifurcation. This analysis also reveals that some of the
transitions observed in the zigzag phase diagram are governed
by these bifurcations; in particular, some part of the $L_2$
transition between defect rich and defect poor dynamics of the
CGLE is apparently given by these bifurcations. It is
therefore unlikely that a unified description of this $L_2$
curve exists.

The existence of parameter regimes where stationary or traveling
zigzags act as building blocks for chaotic and intermittent zigzag
dynamics occurring on very slow scales (see Fig. \ref{f3a} and
\ref{f3b}) illustrates the amazing richness of the one-dimensional
CGLE.

It is a pleasure to acknowledge illuminating discussions with M.\
Howard. MvH acknowledges financial support from the EU under contract
nr. ERBFMBICT 972554.  MI acknowledges financial support from the
Alexander von Humboldt Stiftung, Germany.

\begin{appendix}

\section{appendix}
\label{app}
\subsection{Numerical defect detection}
\label{def}

\begin{figure}
  \vspace{-0.0cm} \epsfxsize=1.0\hsize \mbox{\hspace*{-.04 \hsize}%
  \epsffile{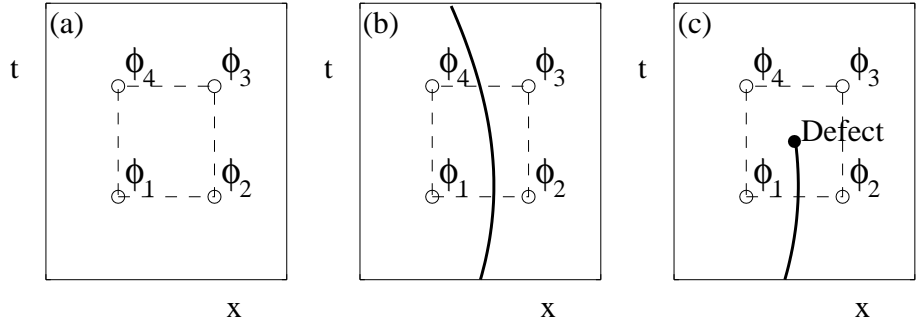} } \vspace{-5.3cm} \caption{Illustration
  of our defect detection algorithm. The open circles
  represent four points of our discretized lattice, and the
  dashed line indicates the loop integral. (a) Regular case,
  no branch lines. (b) When no defects are present, a branch
  line intersects the loop twice.  (c) A defect within the
  loop gives rise to a single intersection with the
  branch line. For further details see text. } \label{defdet}
\end{figure}

Detecting defects in a phase field defined on a discrete
space-time grid is not completely trivial since the phase
variable $\phi$ is defined modulo 2$\pi$ only. Even for a
smooth phase field, $\phi$ jumps by 2$\pi$ along branch lines,
and it is difficult to distinguish between such branch lines
and large ``physical'' gradients of $\phi$. A simple and
robust method to detect defects is illustrated in
Fig.~\ref{defdet}.
If we assume that the discretization is fine enough then the
phase differences between neighboring points should also be
small. Hence we require that $|\phi_2-\phi_1|$ is less than
$\pi$. If we find a larger difference, we assume that this is
because a branch line crosses between these two points (as in
Fig.~\ref{defdet}b), and we simply add or subtract a
correction of $2\pi$ to this difference. Obviously, a branch
line will cut twice through the loop shown in
Fig.~\ref{defdet}b; the two corrections cancel, and our loop
integral will be zero. However, when a defect is present
within this loop, the branch line emanating from this defect
will intersect the loop only {\em once} (Fig.~\ref{defdet}c),
and the addition of the corrections along this loop yields
that the loop integral is plus or minus $2\pi$.  A stable and
fast algorithm to detect defects is thus to mark the bonds in
our space-time lattice where absolute values of the phase
difference between adjacent points are larger then $\pi$ by
$\pm 1$, and then perform the loop integrals over these bond
variables.

\subsection{Numerical integration}

To be able to explore the CGLE for large domains and
integration times, we have chosen a simple next-neighbor,
finite difference scheme, where the resulting set of ODEs are
integrated using an explicit 4th. order Runge-Kutta scheme
with an adaptive time step \cite{NumRecC}. We have taken a
spatial resolution of $\Delta x = 1$ and the time step remains
smaller than 0.05 in general. Clearly, such a code sacrifices
accuracy for speed; we have no indications, however, that the
qualitative zigzag behavior is very sensitive to this. The
stability analysis described in section~\ref{spec} uses the
same integrator to facilitate direct comparison between CGLE
behavior and the linear stability of the zigzags, and a more
refined spatial grid would increase the number of degrees of
freedom used in the stability analysis beyond what we can
handle numerically\footnote{More refined methods, such as
multiple shooting and orthogonal collocation strategies for
ODEs, as well as implementation of an effective Newton-Picard
method \cite{LuRo98,LuRo2000} for the Newton iteration part of
the shooting problem could prove useful; both for more
numerically ill-behaved problems and if a finer discretization
of the CGLE is considered.}.

\subsection{Linear stability calculations}

By discretization of space the linear stability problem for zigzags
 has been converted into a shooting problem for the corresponding
 ODEs.
%
%
To obtain a well-defined problem, we add three pinning equations for
the three unknowns corresponding to the period $T$, the global phase
shift $\phi$ and the velocity $v$.
%
%
If the one-dimensional spatial domain is discretized into $n$
equidistant grid points, after integrating for a period $T$
we obtain $2n+3$ real equations in $2n+3$ real unknowns which
may be solved by a standard Newton iteration procedure.
Therefore the corresponding Jacobian must be be evaluated.  An
accurate numerical determination of the Jacobian matrix
can be found by solving the variational equations associated with the
discretized CGLE for the respective variables and parameters
\cite{Marek83}.
For the numerical integration of these variational equations, we used
the explicit solver from the LSODE package~\cite{Lsode}, and then
linked the solver directly to a standard continuation
package~\cite{Marek91}, which allowed the parametric dependence to be
determined.
Note that for a spatial domain discretized into $n$ grid points, the
integration of the variational equations requires solving a system of
$2n(2n + 1)$ coupled ODEs, which for the problem discussed here
typically is of the order $40000$!

\begin{figure}
  \vspace{-0cm} \epsfxsize=1.\hsize \mbox{\hspace*{-.01 \hsize}
    \epsffile{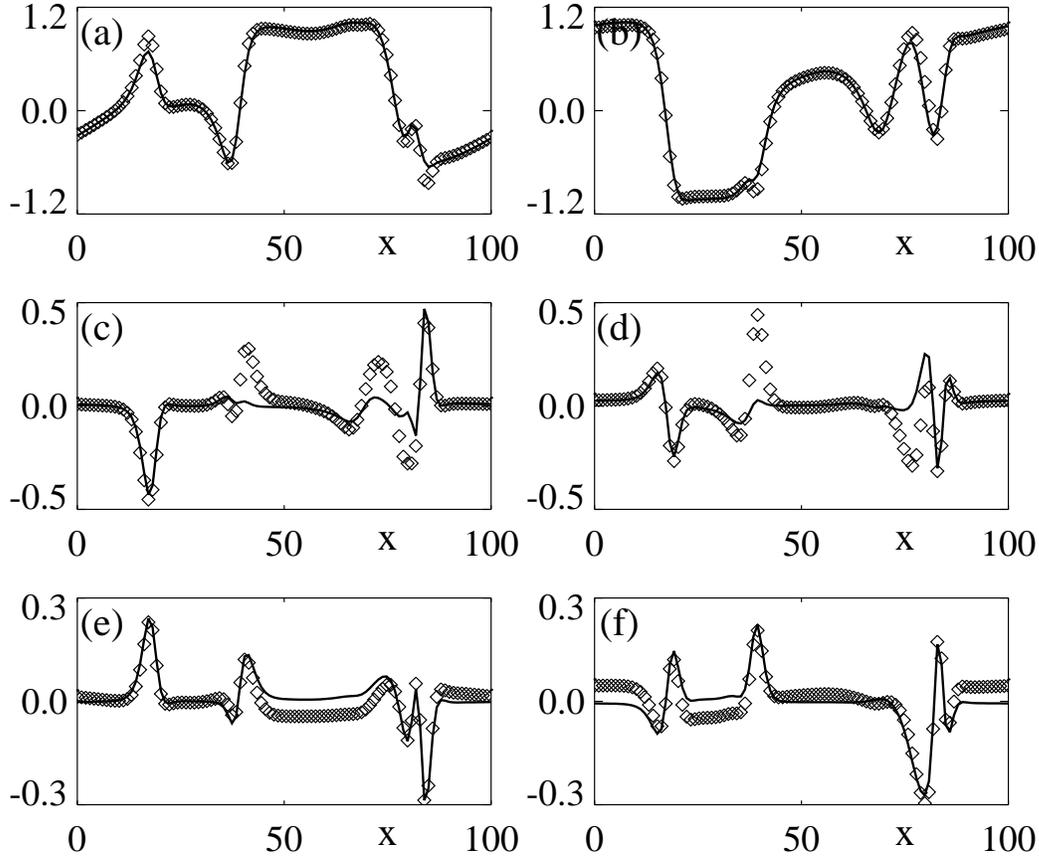} } \vspace{-0cm} \caption{Comparison between
    the real and imaginary parts of the zigzag eigenmodes associated
    with the phase invariance (a-b), translational invariance of time
    and space (c-d) and (e-f), for the CGLE for $c_1 = 0.9$ and $c_3 =
    1.15$. Numerical obtained modes are indicated by symbols, while
    the predicted modes are given by closed curves.} \label{fig:Modes}
\end{figure}

The local stability properties of the periodic solutions thus obtained
are described by Floquet theory~\cite{IJ90}. The time dependent
variation of local perturbations of the limit cycle after the
evolution of each period along the orbit is governed by the monodromy
matrix $M$, which can be obtained as a by-product from the solution of
the variational equations after each successful convergence of a
continuation step~\cite{Marek83}.  The eigenvalues of the monodromy
matrix are the Floquet multipliers and describe the growth or decay of
perturbations of our limit cycle; loosely they are referred to as the
``spectrum'' of the corresponding zigzags.  For a given set of
parameters, the condition $\abs{\lambda} = 1$, designate bifurcation
points where zigzags change stability.

%


For limit cycle solutions of systems of autonomous ODEs, one
multiplier will always satisfy $\lambda = 1$ due to the time
translational invariance of the periodic orbit.  Additional
invariances appear for the problem considered here. Due to
phase invariance $\I A$ is a neutral eigenmode associated with
a multiplier $\lambda = 1$.  Furthermore, due to the periodic
boundary conditions, the spatial translational mode given by
the gradient $\partial_x A$ of the complex field also
corresponds to a neutral mode.  We should therefore have three
neutral modes, and this provides a convenient check for the
numerical accuracy of the calculated monodromy matrix.

Here we discuss the case $c_1 = 0.9$ and $c_3 = 1.15$,
corresponding to the zigzag pattern shown in~\fig{fig:TestZZ}.
The respective neutral Floquet multipliers associated with the
described modes, differ from unity by \sci{1.706}{-4},
\sci{1.905}{-6}, \sci{9.559}{-4} for time, space, and phase
invariant modes respectively.  Since the eigenmodes associated
with these particular multipliers are known, we may compare
these with the corresponding eigenmodes obtained numerically
from the monodromy matrix. In \figa{fig:Modes}{a-f} we compare
the numerically calculated neutral eigenmodes with the
predictions; the agreement is quite close. The main deviations
between the calculation and prediction are observed in the
spatial translation mode. This is likely due to the rather
coarse discretization used in our numerics ($\Delta x = 1$).

\end{appendix}

\bibliographystyle{elsart-num}
\bibliography{literature} 
 
\end{document}